\documentclass[aps,
prl,
reprint,
nofootinbib,
superscriptaddress,
longbibliography
]{revtex4-2}

\usepackage[centertags]{amsmath}
\usepackage{amssymb}
\usepackage{latexsym}
\usepackage{enumerate}
\usepackage{graphicx}
\usepackage{mathrsfs}
\usepackage[hidelinks]{hyperref}
\usepackage{stmaryrd}
\usepackage{import}
\usepackage{tensor}
\usepackage{xcolor}
\usepackage{bm}
\usepackage{multirow}
\usepackage{breakurl}
\usepackage{float}
\usepackage{siunitx}
\usepackage{mathtools}
\usepackage{autobreak}

\usepackage{orcidlink}
\mathtoolsset{centercolon}

\usepackage{graphicx}
\usepackage{subcaption}

\newcommand{\asinh}{{{\rm{arcsinh}}}}

\allowdisplaybreaks

\begin{document}

\title{Analytical Time-Domain Waveforms for Black-Hole Scattering\\ in the Self-Force Expansion}

\begin{abstract}
    We obtain a systematic understanding of the time-domain gravitational waveform of hyperbolic black-hole scattering at first order in the self-force expansion and all orders in the post-Newtonian/post-Minkowskian (PN/PM) expansion. The continuous frequency spectrum of unbound orbits makes the inverse Fourier transform to the time domain diverge order by order; we introduce a dimensional-regularization-inspired prescription that cures this uniformly. We show that the full waveform then follows, at any PM and PN order, from a single family of integrals, generated by the Lauricella function $F_D^{(3)}$, which we calculate using strategies familiar from the study of Feynman integrals. As a proof of principle, we use this understanding to compute the quadrupolar 1SF waveform through relative 10PM--20PN, that is to order ${\cal O}(G^{10} v^{42})$, the full 1SF waveform to 10PM--10PN, as well as the asymptotic fluxes and the dissipative scattering angle at the corresponding accuracy.
\end{abstract}

\author{Chris Kavanagh\,\orcidlink{0000-0002-2874-9780}}
\affiliation{School of Mathematics \& Statistics, University College Dublin, Belfield, Dublin 4, Ireland, D04 V1W8}

\author{Julio Parra-Martinez\,\orcidlink{0000-0003-0178-1569}}
\affiliation{Institut des Hautes Etudes Scientifiques, 91440 Bures-sur-Yvette, France}

\author{Davide Usseglio\orcidlink{0000-0003-2427-9547}}
\affiliation{Institut des Hautes Etudes Scientifiques, 91440 Bures-sur-Yvette, France}

\maketitle

\noindent \textbf{Introduction}.
The detection of gravitational waves from compact binary coalescences \cite{LIGOScientific:2016aoc,LIGOScientific:2017vwq} has opened a new observational window on strong-field gravity. Present and future observatories \cite{LISA:2017pwj,Punturo:2010zz} rely on waveform models of increasing analytical accuracy. Hyperbolic encounters between black holes contribute to this program in two ways. Scattering observables provide theoretical input for the bound-orbit models used in detection, through effective-one-body \cite{Buonanno:1998gg,Buonanno:2000ef,Damour:2016gwp,Pompili:2023tna,Nagar:2023zxh,Gamboa:2026jht} and boundary-to-bound maps \cite{Kalin:2019rwq,Kalin:2019inp,Cho:2021arx}, and they are the setting in which the analytic approaches to the relativistic two-body problem can be compared most directly \cite{Kovacs:1977uw,Kovacs:1978eu,Goldberger:2004jt,Barack:2009ux,blanchet2024postnewtoniantheorygravitationalwaves,Damour:2017zjx,Levi:2018nxp,Kosower:2018adc,Antonelli:2019ytb,Bern:2019nnu,Kalin:2020mvi,Damour:2020tta,Bern:2021dqo,DiVecchia:2021ndb,Herrmann:2021lqe,Jakobsen:2021smu,Pound:2021qin,Mougiakakos:2021ckm,DiVecchia:2021bdo,Herrmann:2021tct,Long:2021ufh,Dlapa:2021npj,Jakobsen:2021lvp,Cristofoli:2021vyo,Bern:2021yeh,Bjerrum-Bohr:2022blt,Khalil:2022ylj,Barack:2022pde,Dlapa:2022lmu,Damour:2022ybd,Brandhuber:2023hhy,Elkhidir:2023dco,Georgoudis:2023lgf,Herderschee:2023fxh,Barack:2023oqp,Whittall:2023xjp,Jakobsen:2023ndj,Rettegno:2023ghr,Caron-Huot:2023vxl,Jakobsen:2023hig,Bini:2023fiz,DeAngelis:2023lvf,Georgoudis:2023eke,Bini:2024rsy,Buonanno:2024vkx,Brunello:2024ibk,Driesse:2024xad,Buonanno:2024byg,Long:2024ltn,Bini:2024ijq,Biswas:2024ept,Driesse:2024feo,Damour:2025uka,Ivanov:2025ozg,Cipriani:2026xmx,Brunello:2025eso,Warburton:2025ymy,Bern:2025wyd,Driesse:2026qiz,Barack:2026izc,Bini:2026dvn,DeAngelis:2026ymn}. These approaches include the post-Newtonian (PN) expansion, in powers of $v^2/c^2\sim GM/(rc^2)$ for slowly moving and weakly bound bodies, the post-Minkowskian (PM) expansion, in powers of $GM/(rc^2)$ at arbitrary velocity, effective field theory, scattering amplitudes and the gravitational self-force (GSF). In addition, close flybys and dynamical captures in dense stellar environments are candidate burst sources for current and future detectors \cite{Kocsis:2006hq,Mukherjee:2020hnm,Codazzo:2022aqj}.

When one black hole is much lighter than the other, a prime target of space-based detectors~\cite{LISA:2017pwj}, the natural framework is GSF theory. At first order in the mass ratio (1SF), the lighter body follows a geodesic of the heavier black hole and sources a linear perturbation of its geometry. This approximation holds at large velocities and in the strong-field regime, and it has been used to compute gravitational observables to very high PN order for bound orbits and to high combined PM--PN order for hyperbolic encounters~\cite{Kavanagh:2015lva,Castillo:2024isq,Munna:2022gio,Munna:2022xts,Geralico:2025rof,Geralico:2026kbm,Geralico:2026efi}. Scattering waveforms at 1SF are known to high order in the frequency domain~\cite{Geralico:2026kbm,Geralico:2026efi}, and the first time-domain results have appeared recently~\cite{Geralico:2026jit, DeAngelis:2026kew}. These results are obtained order by order, and a description of the time-domain waveform valid at arbitrary order has been missing. 

In this Letter we give an all-orders description of the time-domain scattering waveform at 1SF. We determine its structure at any PM and PN order and for any multipole. The main challenge for this computation is specific to unbound motion. A bound binary radiates at discrete frequencies, whereas a scattering orbit radiates a continuous spectrum, and the inverse Fourier transform to the time domain produces divergent integrals at each order of the PM--PN expansion, even though their sum is finite. We overcome this challenge by regulating the divergent integrals with a prescription borrowed from dimensional regularization~\cite{Damour:2001bu,Blanchet:2003gy}. We multiply the frequency integrand by a factor $\bar\omega^{-\varepsilon}$, which makes each integral finite and acts in the same way on instantaneous and hereditary contributions. Second, we show that, once regulated, the integrals at all orders belong to one family, generated by the Lauricella function $F_D^{(3)}$~\cite{Exton1996}, a three-variable generalization of the Gauss hypergeometric function. We evaluate this family with methods developed for Feynman integrals, namely integration-by-parts reduction to master integrals and differential equations. The result fixes the functions that appear in the waveform at any order: harmonic polylogarithms whose transcendental weight grows by one unit per PM order, as in the loop-by-loop growth of transcendentality in scattering amplitudes.

As a demonstration of our strategy, we compute the quadrupolar ($l=2$) waveform through relative 10PM--20PN order and the full 1SF waveform through 10PM--10PN, which requires the multipoles $l=2,\dots,22$. From these we derive the associated asymptotic energy and angular momentum losses, together with the dissipative corrections to the scattering angle at 13PM-10PN, going multiple orders beyond the state of the art in both expansions.

\noindent \textbf{From black-hole perturbation theory to a generating family of integrals}.\label{sec2} We consider a point particle of mass $\mu$ on an equatorial scattering geodesic of a Schwarzschild black hole of mass $M$, parametrized by the impact parameter $b$ and the asymptotic velocity $v$ ($0<v<1$), in units $G=c=1$. The particle sources a perturbation governed by the $s=-2$ Teukolsky equation, which we solve mode by mode in the frequency domain by the standard algorithm \cite{Sasaki:2003xr,Pound:2021qin}: the homogeneous solutions are constructed analytically with Mano-Suzuki-Takasugi (MST) techniques \cite{Mano:1996vt}, the point-particle source is projected onto spin-weighted spherical harmonics ${}_{-2}Y_{lm}(\theta,\phi)$, and the convolution of source and homogeneous solution yields the asymptotic Weyl scalar $\psi_4$, related to the strain by $r^4\psi_4\sim-\tfrac{1}{2}\ddot{h}$, with $h=h_+-ih_\times$ and the dot denoting a derivative with respect to the retarded time $u:=t-r_*$. This pipeline is standard in the analytical self-force treatment of both bound \cite{Bini:2013rfa,Kavanagh:2015lva,Munna:2022gio} and unbound \cite{Bini:2024icd,Geralico:2025rof} orbits. Integrating twice to reconstruct $h$ introduces the homogeneous freedom $h\to h+C_0+C_1u$, so the retarded/no-incoming boundary conditions must be imposed consistently, together with the choice of zero initial strain, rather than assumed to transfer automatically from $\psi_4$.

Analytical waveforms require a perturbative regime: as in \cite{Bini:2024icd}, we adopt the double expansion $b/M\gg1$ (PM) and $v\ll1$ (PN), and introduce the rescaled variables
\begin{equation}
    \bar{t} = vt/b\,, \qquad\bar{u}=vu/b \,, \qquad \bar{\omega} = b\omega/v\,.
\end{equation}
In these variables the radial function $R_{lm\omega}$, which collects the MST homogeneous solution and the source coefficients evaluated along the orbit $r_p(t)$, admits the expansion
\begin{equation}
R_{lm\omega}({r}_p(t))=\sum_{j,k}f^{(j,k)}_{lm}(\bar t)\,(-i\bar{\omega})^j\log^k(-i\bar{\omega})\,,
\end{equation}
with the ranges of $j,k$ set by the target PM and PN orders (see e.g. \cite{Neef:2026qoq} for an open source algorithm generating the homogeneous expansion), and the functions $f^{(j,k)}_{lm}$ encode all the dependence on the geodesic source. The $(l,m)$ mode of the strain then takes the double-integral form
\begin{align}\nonumber
    h_{lm}(\bar{u},r) & = \dfrac{2\mu}{r}\sum_{j,k} \int \dfrac{d\bar{\omega}}{2\pi} \int d\bar{t} \, e^{i\bar{\omega}(\bar{t}-\bar{u})}\,(-i\bar{\omega})^{j-2}  \\
    & \times \log^k(-i\bar{\omega})\,  f^{(j,k)}_{lm}(\bar{t})\,.\label{eq:WF_generic}
\end{align}

\paragraph*{Divergences and regularization.}
The frequency dependence of each term in Eq.~\eqref{eq:WF_generic} is a power times a logarithm, so the natural strategy is to exchange the two integrations and perform the frequency integral first. This exchange is not guaranteed \textit{a priori}. For bound orbits it is harmless: the frequency integral collapses to a sum over harmonics of the orbital frequencies, and the time-domain waveform is reconstructed without issue. For unbound motion a genuine Fourier integral remains, and the exchange produces divergent expressions term by term, even though their sum is finite.

In \cite{Bini:2024icd} the exchange was justified with an exponential regulator. Here instead, 
inspired by dimensional regularization, in which integrals are regulated by analytic continuation to $D=4-2\varepsilon$ dimensions~\cite{Damour:2001bu,Blanchet:2003gy}, we regulate these divergences by multiplying the frequency integrand in Eq.~\eqref{eq:WF_generic} by $(-i\bar\omega)^{-\varepsilon}$, which defines a regulated waveform $h^{\varepsilon}_{lm}$. For suitable complex $\varepsilon$ all integrals converge and the integrations can be exchanged, producing at most poles in $\varepsilon$ term-by-term, which cancel in the sum. The physical waveform is recovered at the end of the computation as $h_{lm}=\lim_{\varepsilon\to0}h^{\varepsilon}_{lm}$. 

The regulator also generates the logarithms in $\bar\omega$, through the identity
\begin{equation}\label{eq:log_eps}
    (-i\bar\omega)^{j-2-\varepsilon}\log^k(-i\bar\omega)=(-\partial_\varepsilon)^{k}(-i\bar\omega)^{j-2-\varepsilon}\,.
\end{equation}
The frequency integral therefore reduces to the elementary Fourier transform of a power $\alpha$ of the frequency,
\begin{equation}
   \int_{-\infty}^{\infty} \dfrac{d\bar{\omega}}{2\pi}\, e^{i\bar{\omega}(\bar{t}-\bar{u})}\, (-i\bar\omega)^{\alpha} = \dfrac{\theta(\bar{u}-\bar{t})}{\Gamma(-\alpha)\,(\bar{u}-\bar{t})^{\,\alpha+1}}\,,
\end{equation}here with $\alpha=j-2-\varepsilon$, followed by $k$ derivatives in $\varepsilon$. The Heaviside function restricts the time integral to $\bar t<\bar u$: each mode becomes a convolution over the past history of the source against the power-law kernel $(\bar u-\bar t)^{\,1-j+\varepsilon}$. 

Note that, in contrast with the exponential regulator of \cite{Bini:2024icd}, the dimensional regulator acts uniformly on all sectors.
For every $k$, including $k=0$, the exponent of the kernel depends on $\varepsilon$, so the frequency integral cannot be identified with a delta function or its derivatives: formally no local-in-time contribution survives, and the regularized waveform is fully non-local. We will now explain that this allows a single generating family of integrals to capture the complete waveform.

\paragraph*{A generating family for the waveform.}

Working at fixed $(l,m)$ and $(j,k)$, and leaving these indices implicit from now on, the regulated waveform is therefore a single integral over the power-law kernel,
\begin{equation}\label{eq:wf_mode_dimreg}
h^\varepsilon(\bar{u},\bar{r}) \supset \dfrac{2\mu}{r}\,(-\partial_\varepsilon)^{k}\!\int_{-\infty}^{\bar u}\! d\bar{t}\,\dfrac{f(\bar{t})}{\Gamma(2-j+\varepsilon)\,(\bar u-\bar t)^{\,j-1-\varepsilon}}\,,
\end{equation}
where $f(\bar{t})\equiv f^{(j,k)}_{lm}(\bar{t})$. Because $f$ descends directly from the source coefficients, its generic structure is easily determined:
\begin{equation}
    f(\bar{t}) \supset \dfrac{\bar{t}^{\,a_1}\,{\rm arcsinh}^{p_1}(\bar{t})\,\arctan^{p_2}(\bar{t})\,\log^{p_3}(1+\bar{t}^2)}{(1+\bar{t}^2)^{a_2/2}}\,,
\end{equation}
with positive integer powers $a_i,p_i$. This form is universal at 1SF: the perturbative  PM--PN order, through $j$ and $k$, merely fixes which powers contribute.

Despite its apparent complexity, the remaining time integral can be brought to a universal form. With the change of variables
\begin{equation}
    \bar{u} = \dfrac{y^2-1}{2y}\,,\qquad \bar{t} = \dfrac{y'^2-1}{2y'}\,,
\label{eq:y_variable}
\end{equation}
the time integral runs from $0$ to $y$, and its singular points are isolated at $\{0,\pm i, y, -1/y\}$. In this variable the inverse trigonometric and hyperbolic functions in $f$ become logarithms of rational functions of $y'$,
\begin{align}\nonumber
    \asinh\bar t&=\log y'\,,\qquad
    \arctan\bar t=i\log\!\Big(\tfrac{i(1-iy')}{1+iy'}\Big)\,,\\
    \log(1+\bar t^2)&=2\log\!\Big(\tfrac{(1+iy')(1-iy')}{2y'}\Big)\,.
\end{align}
Summing over their powers with three generating variables $\{x_1,x_2,x_3\}$ turns these logarithms into powers of $y'$, $1+iy'$ and $1-iy'$,
\begin{align}\nonumber
    &\sum_{p_i}\frac{x_1^{p_1}(-i x_2)^{p_2}(x_3/2)^{p_3}}{p_1!\,p_2!\,p_3!}\,
    \asinh^{p_1}\bar t\,\arctan^{p_2}\bar t\\ \nonumber
    &\hspace{2.5cm}\times\log^{p_3}(1+\bar t^2)\\
    &=i^{x_2}\,2^{-x_3}\,y'^{\,x_1-x_3}\,(1+iy')^{\,x_3-x_2}\,(1-iy')^{\,x_2+x_3}\,,\label{eq:source_GF}
\end{align}
so that each source function can be recovered by expanding this generating function. Collecting everything, the waveform mode can be expressed in terms of a single integral family, which we recognize as the Euler-type integral representation of the Lauricella function $F_D^{(3)}$
\par\medskip\noindent
{\setlength{\fboxsep}{2pt}\medmuskip=2mu
\fbox{\begin{minipage}{\dimexpr\linewidth-2\fboxsep-2\fboxrule\relax}
\setlength{\abovedisplayskip}{0pt}\setlength{\belowdisplayskip}{0pt}
\setlength{\abovedisplayshortskip}{0pt}\setlength{\belowdisplayshortskip}{0pt}
\begin{align}
    &\int_{0}^y\! dy'\,  (y-y')^{\lambda_1-1} y'^{\lambda_2-1}(1+y y')^{\lambda_3}(1+iy')^{\lambda_4}(1-iy')^{\lambda_5}\nonumber\\
    &= y^{\lambda_1+\lambda_2-1}\, \mathrm{\beta}(\lambda_1,\lambda_2)\label{eq:GF_main}\\
    &\hspace{0.4cm}\times F^{(3)}_D\!\big(\lambda_2;-\lambda_3,-\lambda_4,-\lambda_5;\lambda_1+\lambda_2;-y^2,-iy,iy\big)\nonumber
\end{align}
\end{minipage}}}
\par\medskip\noindent
where $\mathrm{\beta}$ is the Euler Beta function. The exponents $\lambda_i=\nu_i+\alpha_i$ split into integer parts $\nu_i$, fixed by the source and the perturbative order, and non-integer shifts $\alpha_i$, linear in the regulator and the generating variables,
\begin{align}\nonumber
       \boldsymbol\alpha&=\big(\varepsilon,\;x_1-x_3-\varepsilon,\;\varepsilon,\;x_3-x_2,\;x_2+x_3\big)\,.
\end{align}

This is the main analytical result of this Letter. Eq.~\eqref{eq:GF_main} provides a generating function for all of the integrals in the scattering waveform, valid for any multipole and at any order in the PM--PN expansion at 1SF. The expansion of the Lauricella function around integer arguments reproduces the entire logarithmic structure characteristic of scattering observables. This expansion can be easily carried out using methods familiar from the study of Feynman integrals: integration-by-parts identities reduce the integral family from arbitrary integer shifts to four master integrals, which satisfy a system of differential equations in $y$. Solving this system order by order in the regulators expresses the waveform in terms of harmonic polylogarithms. This is explained in more detail in the Supplemental Material.

\noindent\textbf{Results}.\label{sec5}
\begin{figure*}[t]
  \centering
  \captionsetup[subfigure]{font=small,margin={30pt,0pt}}

  \begin{subfigure}[c]{0.42\linewidth}
    \centering
    \includegraphics[width=\linewidth]{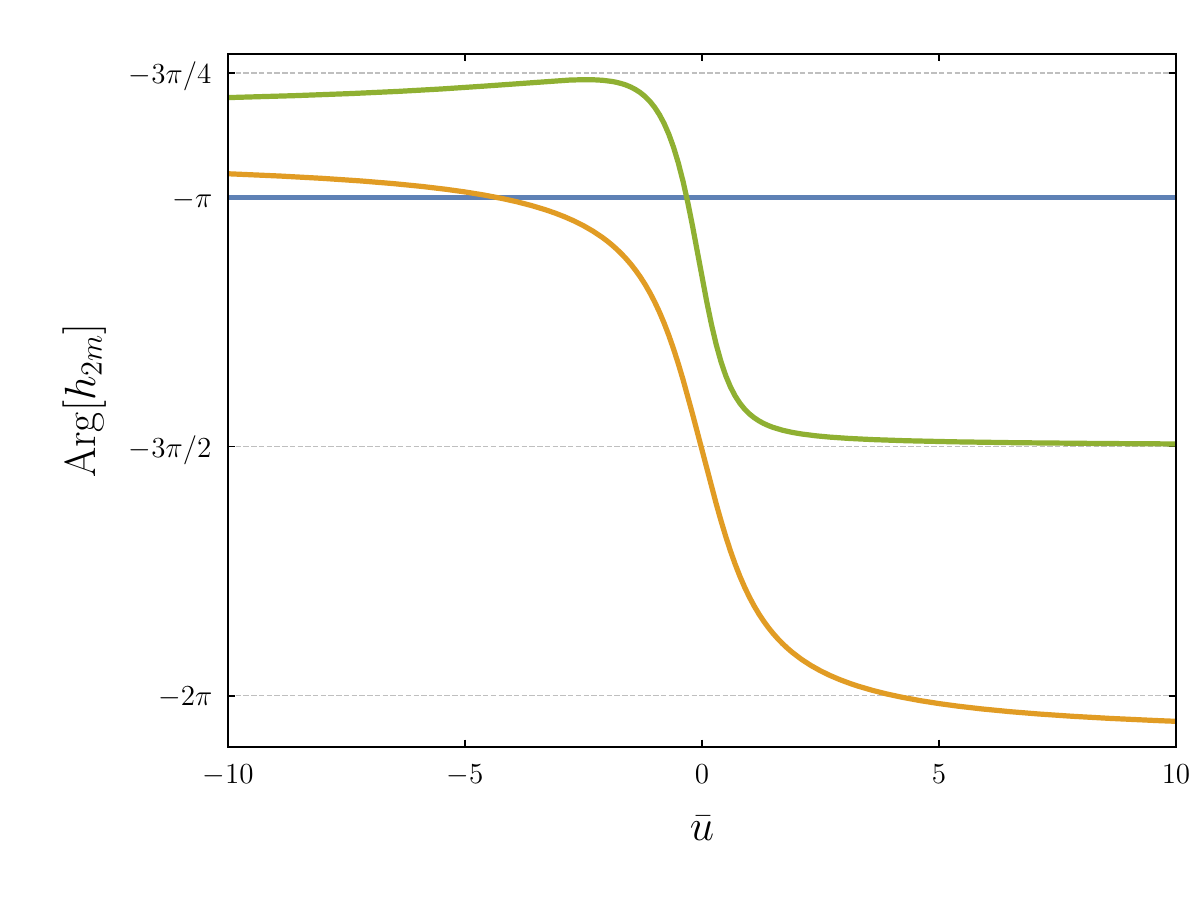}
    \caption{Phase (unwrapped).}
    \label{fig:waveform-phase}
  \end{subfigure}\begin{minipage}[c]{0.16\linewidth}
    \centering
    \raisebox{12pt}{\includegraphics[width=0.85\linewidth,trim=25 205 28 162,clip]{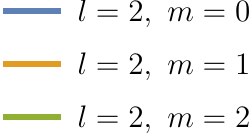}}
  \end{minipage}\begin{subfigure}[c]{0.42\linewidth}
    \centering
    \includegraphics[width=\linewidth]{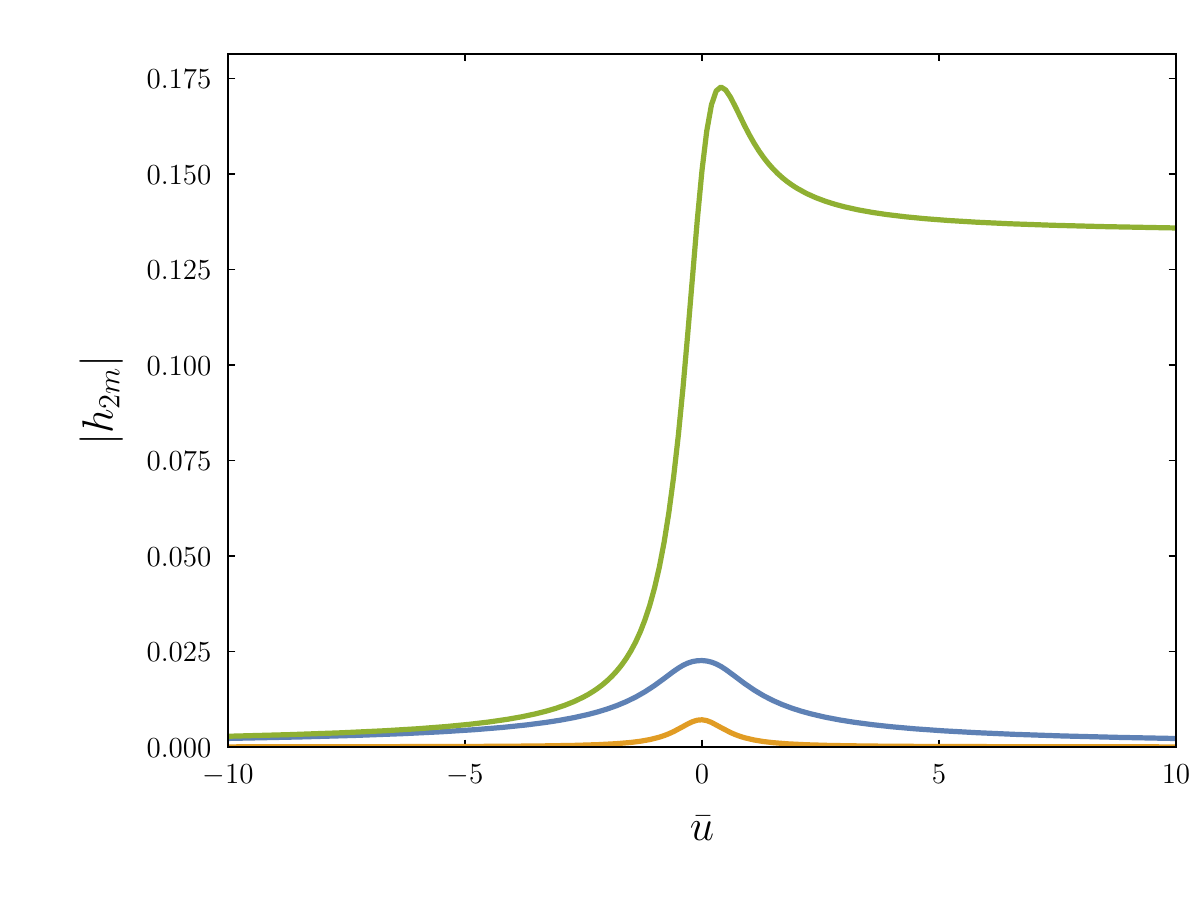}
    \caption{Amplitude.}
    \label{fig:waveform-amplitude}
  \end{subfigure}

\caption{Quadrupolar modes evaluated through 10PM-20PN at $v=0.2$,
    $b=100$, and $\epsilon=1/(bv^2)=0.25$.}
  \label{fig:waveform-modes}
\end{figure*}
We have implemented the strategy outlined above in a custom algorithm. As a concrete application of our formalism, we have computed at 1SF:
\begin{itemize}
    \item the quadrupolar $l=2$ waveform to 10PM-20PN,
    \item the full waveform to 10PM-10PN,
    \item the fluxes of energy and angular momentum to 12PM-10PN,
    \item the dissipative scattering angle to 13PM-10PN,
\end{itemize}
all of which are well beyond the state of the art.
As a reference benchmark, the full calculation of the quadrupolar waveform takes 23 hours on an AMD EPYC node with 48 cores and 128 GB of RAM.

The complete results are too large to be printed here, so we have made them available in machine-readable form in a GitHub repository \cite{ancillary_file}, which contains the 10PM-10PN waveform with all modes $l=2,\dots,22$ and the 10PM-20PN quadrupolar waveform. In addition, an ancillary file to this manuscript contains the 9PM-9PN quadrupolar waveform, the 12PM-10PN asymptotic fluxes, and the 13PM-10PN dissipative scattering angle. We now describe some features of each of these results and the checks that we have carried out on them.

\textit{The quadrupolar waveform.}
The $l=2$ modes depend on the retarded time through the variable $y$ of Eq.~\eqref{eq:y_variable}, and on $v$ and $b$ through the PM--PN expansion. We organize each mode as a series in $M/(bv^2)$,
\begin{equation}
        h_{22}(y,b)=\dfrac{\mu}{r}\sum_{n=1}^{10}\left(\frac{M}{bv^2}\right)^{n}h^{(n)}_{22}(y,b)\,,
\label{eq:h2m_explicit}
\end{equation}
where the $n$PM coefficient $h^{(n)}_{22}$ is itself a series in $v$, which we computed through 20PN. 

The hierarchy of non-local-in-time effects is reflected in the residual dependence of $h^{(n)}$ on $b$ through powers of the logarithm 
\begin{equation}
    \iota(b)=\log\!\left(\frac{8vM}{b}\right).
\end{equation}
Linear terms in $\iota$ (i.e. tail) first appear at 2PM--1.5PN, quadratic terms (i.e. tail-of-tail) at 3PM--3PN, and, more generally, $\iota^n$ first appears at $(n+1)$PM--$(3n/2)$PN, with subleading corrections entering at higher PN and PM orders.

The functions of $y$ in $h^{(n)}_{22}$ can be organized by transcendental weight. We assign weight zero to rational functions, weight one to logarithms and to inverse trigonometric and hyperbolic functions, weight two to dilogarithms and to products of two weight-one functions, and so on, with $\pi$ of weight one and $\zeta(n)$ of weight $n$. With this counting, the $n$PM coefficient contains functions of weight up to $n-1$, so the 10PM term reaches weight nine. To all orders, the functions appearing belong to the general class of harmonic polylogarithms \cite{Remiddi:1999ew}. The PN expansion changes only the rational coefficients multiplying these functions.

As an example we show the 5PM-1.5PN term of the $l=m=2$ mode, which lies beyond the state of the art and contains the first non-local contributions at 5PM:
\begin{equation} h_{22}^{(5)}(y,b)|_{1.5PN}  = \frac{16  y^2 v^5}{3\left(1+y^2\right)^{11}}\sqrt{\frac{\pi }{5}}\sum_{w=0}^{4}\mathcal H^{(w)}(y,b),
\label{eq:weight_decomposition}
\end{equation}
where the functions ${\cal H}$ have transcendental weight $w$. Using the shorthand
\begin{equation}
\begin{aligned}
 &\ell=\log y,\quad \rho=\log(1+y^2),\quad L^{\pm}_n=\operatorname{Li}_n(-y^2)\pm\zeta(n),
\end{aligned}
\label{eq:weight_notation}
\end{equation}
they are given by
\begin{align}
&\mathcal H^{(0)}=A_0,
\notag\\[2pt]
&\mathcal H^{(1)}=A_1\ell-A_2\iota,
\notag\\[2pt]
&\mathcal H^{(2)}
=A_3\bigl(L^{-}_2-\iota\ell+2\rho\ell\bigr)-A_4\ell^2,
\label{eq:explicit_weight_sectors}\\[2pt]
&\mathcal H^{(3)}
=A_5\bigl(\tfrac12\iota\ell^2+\ell L^{+}_2-L^{+}_3\bigr)+A_6\ell^3,
\notag\\[2pt]
&\mathcal H^{(4)}
=A_7\bigl(\ell^4+\iota\ell^3-2\rho\ell^3-3\ell^2L^{-}_2+6\ell L^{-}_3-6L^{-}_4\bigr),
\notag
\end{align}
where the rational coefficients $A_i$ are listed in Table~\ref{tab1}.
\begin{table}[tb]
\centering
\begingroup
\scriptsize
\setlength{\jot}{0pt}

\(\boxed{\begin{aligned}
& A_0 = (y-i)^3(y+i)^7
\bigl(5i+39y+162iy^2+1765y^3-3278iy^4\\
&\qquad {}-1765y^5+162iy^6-39y^7+5iy^8\bigr),
\\[2pt]
& A_1 = 4i(y-i)^2(y+i)^6
\bigl(-3-8iy-134y^2+225iy^3-2057y^4\\
&\qquad {}+2788iy^5+2057y^6+225iy^7
+134y^8-8iy^9+3y^{10}\bigr),
\\[2pt]
& A_2 = 12i(y-i)^3(y+i)^7
\bigl(1+3iy+18y^2-107iy^3-166y^4\\
&\qquad {}+107iy^5+18y^6-3iy^7+y^8\bigr),
\\[2pt]
& A_3 = 48\,y(y-i)^2(y+i)^6
\bigl(1-iy+81y^2-391iy^3-530y^4\\
&\qquad {}+391iy^5+81y^6+iy^7+y^8\bigr),
\\[2pt]
& A_4 = 4\,y(y-i)(y+i)^5
\bigl(12-27iy+1241y^2-5378iy^3-7237y^4\\
&\qquad {}+1634iy^5-3539y^6+4030iy^7
+727y^8-3iy^9+12y^{10}\bigr),
\\[2pt]
& A_5 = 96\,y^2(y-i)(y+i)^5
\bigl(3i-11y+350iy^2+859y^3-1346iy^4\\
&\qquad {}-859y^5+350iy^6+11y^7+3iy^8\bigr),
\\[2pt]
& A_6 = 16i\,y^2(y+i)^4
\bigl(9+25iy+1195y^2-3328iy^3-5832y^4\\
&\qquad {}+4392iy^5+1848y^6+64iy^7
+217y^8-19iy^9+3y^{10}\bigr),
\\[2pt]
& A_7 = 192\,y^3(y+i)^4
\bigl(1+14iy+56y^2-156iy^3-180y^4+156iy^5\\
&\qquad {}+56y^6-14iy^7+y^8\bigr).
\end{aligned}}\)

\par
\endgroup
\caption{Rational coefficients entering the (2,2)-mode
of the waveform at 5PM-1.5PN.}
\label{tab1}
\end{table}

Figure~\ref{fig:waveform-modes} shows the phase and amplitude of the individual $m\ge0$ modes of $l=2$, computed at 10PM-20PN, over the full scattering encounter for $b=100M$ and $v=0.2$. Negative $m$ modes are the complex conjugates of the corresponding positive modes. For all modes, the amplitude vanishes at past null infinity, as required by our boundary conditions (zero initial strain and no incoming radiation), specifically $h_{2m}(\bar{u})\sim\mathcal{O}(1/\bar{u})$ as $\bar{u}\rightarrow-\infty$. At future null infinity it generally approaches a nonzero constant, indicating the presence of a linear memory contribution.

We have compared our results with the leading-PM time-domain waveform of \cite{DeAngelis:2026kew}, by expanding the result in the low-velocity regime to the required order and performing a rotation to align the reference frames. The two results agree after applying a well-known Bondi--Metzner--Sachs (BMS) transformation \cite{Veneziano:2022zwh,Bini:2024rsy,Georgoudis:2024pdz}, which maps our result with no initial strain to the one in \cite{DeAngelis:2026kew}. The same term is crucial for the comparison with \cite{Geralico:2026jit}, whose waveform carries a non-zero initial strain as $\bar{u}\to-\infty$. We also find full agreement on the overlapping terms (i.e., through 4PM-7PN).

In the frequency domain, we compared our results with
\cite{Bini:2024rsy}. After accounting for static contributions
and the phase shift induced by a time translation, as discussed
in \cite{Bini:2026dvn}, we find agreement 
through 3PM order. It would also be interesting to compare with frequency-domain
results of \cite{Geralico:2026efi},
but we have not attempted this, since it is redundant with the checks above.

\textit{Full waveform, fluxes and dissipative scattering angle.} As mentioned above, we have computed the full 10PM-10PN waveform, which requires all the modes up to $l=22$.
We have used this result to compute the integrated energy and angular momentum fluxes up to 12PM-10PN, as well as the dissipative scattering angle at 13PM-10PN through the linear response formula of Ref.~\cite{Bini:2012ji}. The full expressions are stored in the GitHub repository \cite{ancillary_file}. Here we report the coefficients for each quantity at 8PM-2PN, which is beyond the state of the art:
\begin{small}
\begin{align}
&\begin{aligned}
&\delta\chi^{\infty,\mathrm{diss}}_{8\mathrm{PM}}
=\frac{18311\mu M^7\pi}{40b^8v^9}\biggl[1
+\left(\frac{96082\pi^2}{54933}
-\frac{22235\pi^4}{146488}\right)v\\
&\qquad {}+\frac{64568737}{13843116}v^2
+\mathcal{O}(v^3)\biggr].
\end{aligned}\\[3pt]
&\begin{aligned}
&\Delta E^\infty_{8\mathrm{PM}}
=\frac{3104\mu^2M^7}{75b^8v^9}\biggl[1+\frac{16501}{679}v^2\\
&\qquad {}+\left(2+\frac{10640\pi^2}{291}
-\frac{8768\pi^4}{3395}\right)v^3
-\frac{791989}{18333}v^4+\mathcal{O}(v^5)\biggr],
\end{aligned}\\[3pt]
&\begin{aligned}
&\Delta J^{\infty,\mathrm{rad}}_{8\mathrm{PM}}
=\frac{416\mu^2M^7}{25b^7v^{10}}\biggl[1-\frac{23}{21}v^2+\left(2+40\pi^2-\frac{1984\pi^4}{585}\right)v^3\\
&\qquad {}-\left(\frac{11716}{1053}+\frac{135\pi^2}{13}\right)v^4
+\mathcal{O}(v^5)\biggr],
\end{aligned}
\end{align}
\end{small}
We note that this includes only contributions from the flux at infinity. Starting at this order, there is an additional contribution from the fluxes through the horizon, which we have not included here. 
Also, it is important to point out that, because of our choice of boundary conditions, the loss of angular momentum does not contain a static contribution, which needs to be included by a BMS transformation \cite{Veneziano:2022zwh,DiVecchia:2022owy,Bini:2024rsy} in order to compute the full dissipative scattering angle. This is reviewed in the Supplemental Material.

Finally, we validate the integrated energy and angular momentum
fluxes, as well as the dissipative scattering angle, against
existing results in the literature. The energy flux agrees with the known results through 4PM upon expansion in the low-velocity limit
\cite{Herrmann:2021lqe,DiVecchia:2021bdo,Herrmann:2021tct,Bern:2021dqo,Bern:2021yeh,Dlapa:2021npj,Dlapa:2022lmu,Jakobsen:2023ndj,Jakobsen:2023hig}. It further reproduces the 5PM--1SF expression for the energy flux in \cite{Driesse:2024feo}\footnote{As expected, the dissipative scattering angle differs from the WQFT result at 5PM--3PN order, owing to the conservative
tail-of-tails contributions, as explained in Sec.~V of
\cite{Bini:2025vuk}.}, and also agrees with \cite{Geralico:2026kbm,Geralico:2026efi} through 6PM--6PN order.
Upon restoring the static contribution to the angular momentum loss, we also recover agreement for both the angular momentum 
flux \cite{Damour:2020tta,Manohar:2022dea,Heissenberg:2025ocy} and the dissipative scattering angle \cite{Herrmann:2021lqe,DiVecchia:2021bdo,Herrmann:2021tct,Bern:2021dqo,Bern:2021yeh,Dlapa:2021npj,Dlapa:2022lmu,Jakobsen:2023ndj,Jakobsen:2023hig}. Notably, the dissipative scattering angle we computed does not contain the contributions that come from the energy and angular momentum losses at the horizon, which contribute to the scattering angle starting at eighth PM order.

\section{Conclusions}\label{sec6}
We have established an all-orders description of the time-domain waveform for hyperbolic black-hole scattering at 1SF. A prescription inspired by dimensional regularization brings the PM--PN expansion into a single integral family generated by Lauricella functions, Eq.~\eqref{eq:GF_main}, turning the reconstruction of the waveform into a systematic analytical problem. This structure opens the calculation to methods developed for Feynman integrals. Using integration-by-parts identities and differential equations, we computed the full waveform through 10PM--10PN, including multipoles from $l=2$ to $l=22$, and extended the quadrupolar waveform through 10PM--20PN. From these high-precision results, we extracted the asymptotic fluxes and the associated dissipative scattering angle to high accuracy. These results demonstrate the reach of the framework, pushing analytical scattering waveforms multiple orders beyond the state of the art in both perturbative expansions.

The framework provides a starting point to extend these techniques to conservative dynamics, pursuing a unified analytical description of scattering at 1SF across the double PM--PN expansion. A forthcoming paper will derive the regulator directly from the homogeneous perturbation equations in $D=4-2\varepsilon$ dimensions, placing the prescription on a rigorous footing. Together, these developments strengthen the connection between black-hole perturbation theory and scattering-based approaches to gravitational-wave modelling.

\noindent \textbf{Acknowledgments}
We thank Donato Bini, Thibault Damour, Stefano De Angelis, Andrea Geralico, Vincent He, and Jakob Neef for useful discussions, and Niels Warburton for numerical comparisons and discussions. DU acknowledges sponsorship of the Italian Gruppo Nazionale per la Fisica Matematica (GNFM). CK acknowledges support from Research Ireland under Grant number 21/PATH-S/9610.
The work of J.P.-M. is supported by the European Research
Council (ERC) under the European Union’s Horizon 2020 research and innovation programme, grant agreement ERC-StG-101221094 GravitaS.

\textit{AI usage.} Artificial intelligence tools were used to assist with the development and automation of code. All results and their scientific interpretation were independently verified by the authors.

\bibliography{bibliography}

\appendix
\onecolumngrid
\section*{Supplemental Material}

\section{Expansion of Lauricella function around integer arguments}\label{Appendix_C}\label{app:Lauricella-expansion}

In this appendix we elaborate on the expansion of Lauricella functions around integer arguments. Recall the family of integrals in Eq.~\eqref{eq:GF_main} from the main text.
Rescaling $y'=yz$ maps the integration range of Eq.~\eqref{eq:GF_main} to $z\in[0,1]$, yielding the following representation
\begin{align}
 &y^{\lambda_1+\lambda_2-1}\int_0^1 dz\,(1-z)^{\lambda_1-1}z^{\lambda_2-1}(1+y^2z)^{\lambda_3}(1+iyz)^{\lambda_4}(1-iyz)^{\lambda_5}\,.
\end{align}
Note that for $y>0$ and $z\in[0,1]$ the factors $1\pm iyz$ have positive real part, so all powers take their principal branch.
Writing the linear factors
\begin{equation}
 \mathbf L(z;y)=(1-z,\,z,\,1+y^2z,\,1+iyz,\,1-iyz),
\end{equation}
the integrals to be evaluated belong to the family
\begin{equation}
 I_{\boldsymbol n}(y)
 =\int_0^1 dz\,\mathcal R(z;y)\prod_{j=1}^5L_j(z;y)^{-n_j},
 \qquad
 \mathcal R(z;y)=\prod_{j=1}^5L_j(z;y)^{\alpha_j},
 \label{app:shift-family}
\end{equation}
with
\begin{equation}
 \boldsymbol\alpha=
 (\varepsilon,\,x_1-x_3-\varepsilon,\,
 \varepsilon,\,x_3-x_2,\,x_3+x_2), \qquad  \boldsymbol n\in\mathbb Z^5.
 \label{app:alpha-family}
\end{equation} 
Our goal is to expand the integrals in the small parameters $\boldsymbol{\alpha}$ for any $\boldsymbol{n}$.
This can be achieved in two steps:
\begin{enumerate}
    \item We reduce the family from arbitrary integer shifts $\boldsymbol{n}$ to a finite basis of \emph{master integrals} using integration-by-parts (IBP) relations.
    \item We compute the expansion of the master integrals by deriving and solving differential equations.
\end{enumerate}
Let us now elaborate on each of these.

\paragraph*{Reduction to master integrals.}

Integration-by-parts identities for this family of integrals are homogeneous, as the endpoint terms vanish in a domain of convergent regulated integrals, and the identity is then continued analytically in the exponents. The reduction to a basis is implemented as a Griffiths--Dwork reduction~\cite{Griffiths:1969,Dwork:1962,Lairez:2014qha,Lairez:2022zkj,delaCruz:2024xit}, which removes polynomial factors and lowers higher-order poles to simple poles. The candidate masters are the members of the family with single poles,
\begin{equation}
 m_j(y)=\int_0^1 dz\,\mathcal R(z;y)\,\partial_z\log L_j(z;y)
 =(\partial_zL_j)\,I_{\mathbf e_j}(y),
\end{equation}
where $\mathbf e_j$ is the $j$th unit vector and $\partial_z\mathbf L=(-1,1,y^2,iy,-iy)$ does not depend on $z$. In fact, using \begin{equation}
    \partial_z\mathcal R=\mathcal R\sum_j\alpha_j\,\partial_z\log L_j\,,
\end{equation} we find that these are not independent but obey one relation:
\begin{equation}
 0=\int_0^1 dz\,\frac{\partial\mathcal R}{\partial z}
 =\sum_{j=1}^{5}\alpha_j\,m_j.
 \label{app:master-relation}
\end{equation}
Eliminating $m_1$ leaves the four-component basis
\begin{equation}
 \mathbf m(y)=(m_2,\,m_3,\,m_4,\,m_5)^T.
 \label{app:four-masters}
\end{equation}
Consequently, every integral in Eq.~\eqref{app:shift-family} is a linear combination of these four masters with coefficients rational in $y$ and the exponents. Note that the reduction is carried out at generic exponents, specializing to Eq.~\eqref{app:alpha-family} before expanding about $\boldsymbol\alpha=0$.

\paragraph*{Differential equation.}
Next we want to expand the master integrals in the parameters $\boldsymbol{\alpha}$.
Differentiating the family of integrals under the integral sign, with $\partial_y\mathbf L=L_2\,(0,0,2y,i,-i)$, gives
\begin{align}\nonumber
 \frac{d}{dy}I_{\mathbf n}
 &=2y(\alpha_3-n_3)\,I_{\mathbf n-\mathbf e_2+\mathbf e_3}
 +i(\alpha_4-n_4)\,I_{\mathbf n-\mathbf e_2+\mathbf e_4}-i(\alpha_5-n_5)\,I_{\mathbf n-\mathbf e_2+\mathbf e_5}.
\end{align}
Specializing to master integrals (including the derivatives of the prefactors $\partial_zL_j$) and reducing the shifted integrals back to the basis yields a linear ODE
\begin{equation}
 \frac{d\mathbf m}{dy} = \Omega(y) \mathbf m = 
 \left(\frac{\Omega_0}{y}+\frac{\Omega_i}{y-i}
                    +\frac{\Omega_{-i}}{y+i}\right)\mathbf m.
 \label{app:master-DE}
\end{equation}
The residue matrices are linear in the exponents:
\begin{align}
 \Omega_0&=\begin{pmatrix}
 0&2\alpha_3&\alpha_4&\alpha_5\\
 0&-\alpha_{12}-\alpha_{1245}&0&0\\
 0&\alpha_3&-\alpha_{12}&0\\0&\alpha_3&0&-\alpha_{12}
 \end{pmatrix},\,\,\, \Omega_i=\begin{pmatrix}
 0&0&0&0\\
 \alpha_2&\alpha_{134}&0&\alpha_5\\
 \alpha_2&0&\alpha_{134}&\alpha_5\\
 0&0&0&0
 \end{pmatrix},\,\,\,
 \Omega_{-i}=\begin{pmatrix}
 0&0&0&0\\
 \alpha_2&\alpha_{135}&\alpha_4&0\\
 0&0&0&0\\
 \alpha_2&0&\alpha_4&\alpha_{135}
 \end{pmatrix}.
 \label{app:master-residues}
\end{align}
where we use concatenated subscripts, e.g.,  $\alpha_{12}=\alpha_1+\alpha_2$ and $\alpha_{1245}=\alpha_1+\alpha_2+\alpha_4+\alpha_5$. This is the restriction of the Lauricella differential system to the curve of arguments $(-y^2,-iy,iy)$ appearing in Eq.~\eqref{eq:GF_main}. Its logarithmic kernels restrict the solution to the class of harmonic polylogarithms of argument $-iy$ \cite{Remiddi:1999ew}.

To solve the differential equation, we can use the value of the masters at $y=0$ as a boundary condition, which is simply
\begin{equation}
 \mathbf m(0)=\mathcal B_0\,\mathbf b_0,
 \qquad
 \mathbf b_0=(1,0,0,0)^T,
 \qquad
 \mathcal B_0=
 \frac{\Gamma(1+\alpha_1)\Gamma(\alpha_2)}
 {\Gamma(1+\alpha_{12})}.
 \label{app:master-normalization}
\end{equation}
The solution is then constructed as a path-ordered exponential, algebraically, word by word in the alphabet ${0,+i,-i}$:
\begin{equation}
 \mathbf m(y)= P\text{exp}\left(\int_0^y \Omega(t) dt\right)\mathbf m(0) = \mathcal B_0\sum_{w} G(w;y)\,\mathbf v_{w},
 \qquad \mathbf v_{\varnothing}=\mathbf b_0,
 \qquad \mathbf v_{(\xi,w)}=\Omega_{\xi}\mathbf v_{w}.
 \label{app:master-word-solution}
\end{equation}
in terms of the iterated integrals defined by
\begin{equation}
 G(\xi,w;y)=\int_0^y\frac{dt}{t-\xi}G(w;t),
 \qquad G(\varnothing;y)=1,
 \qquad \xi\in\{0,i,-i\}.
\end{equation}
Each coefficient $\mathbf v_{w}$ is homogeneous of degree $|w|$ in $\boldsymbol\alpha$. Moreover, $\Omega_0\mathbf b_0=0$, so words ending in zero have vanishing coefficients and the nonempty iterated integrals occurring here vanish at the base point $y=0$ and have no logarithmic singularity at that location. Equation~\eqref{app:master-word-solution} therefore directly generates the expansion in the $\boldsymbol{\alpha}$ parameters.

\section{Asymptotic fluxes and dissipative scattering angle}
\label{app:fluxes}

We describe here how to compute the fluxes from the high precision 1SF waveform we derived in the main text. In particular, we describe here how to reintroduce the static contribution to the angular momentum loss to get the dissipative scattering angle. The integrated losses are
  \begin{align}
    \Delta E^\infty
    &=
    \frac{v}{8\pi b}
    \sum_{l,m}
    \int_0^\infty
    \frac{y^2}{1+y^2}
    \left|\frac{dh_{lm}}{dy}\right|^2\,dy,
    \label{eq:SM_energy_loss}
    \\
    \Delta J^{\infty,\mathrm{rad}}
    &=
    \frac{1}{16\pi}
    \sum_{l,m}m
    \int_0^\infty
    \operatorname{Im}\!\left[
      h_{lm}\frac{d\overline{h}_{lm}}{dy}
    \right]\,dy.
    \label{eq:SM_angular_momentum_loss}
  \end{align}
  The overline denotes complex conjugation, and
  $\mathrm{rad}$ distinguishes the radiative angular-momentum
  loss from the static contribution.

  The dissipative 1SF scattering angle follows from the
  linear-response formula~\cite{Bini:2012ji},
  \begin{equation}
    \delta\chi^{\infty,\mathrm{diss}}
    =
    -\frac12
    \left[
      \left(\frac{\partial\chi_{\mathrm{cons}}}{\partial E}\right)_J
      \Delta E^\infty
      +
      \left(\frac{\partial\chi_{\mathrm{cons}}}{\partial J}\right)_E
      \Delta J^\infty
    \right],
    \label{eq:SM_dissipative_scattering_angle}
  \end{equation}
  where $\chi_{\mathrm{cons}}(E,J)$ is the geodesic scattering
  angle. The derivatives are evaluated at the incoming physical
  energy and angular momentum, with $M$ and $\mu$ held fixed.
  The angular-momentum loss entering this relation includes
  both the radiative and static contributions and is defined
  in the incoming center-of-mass frame.

  Our waveform has zero initial strain. Restoring the constant
  incoming strain $h^-_{lm}$ gives
  \begin{equation}
    h^{\mathrm{full}}_{lm}(y)=h_{lm}(y)+h^-_{lm}.
  \end{equation}
  This leaves the energy loss unchanged but adds the boundary
  contribution
  \begin{equation}
    \Delta J^{\infty,\mathrm{static}}
    =
    \frac{1}{16\pi}
    \sum_{l,m}m\,
    \operatorname{Im}\!\left[
      h^-_{lm}\,\Delta \overline{h}_{lm}
    \right],
    \label{eq:SM_static_angular_momentum_loss}
  \end{equation}
  where $\Delta h_{lm}=h^+_{lm}-h^-_{lm}$ is the waveform
  memory and $h^\pm_{lm}$ are the endpoints of the restored
  waveform. This expression follows directly by substituting
  $h^{\mathrm{full}}_{lm}$ into
  Eq.~\eqref{eq:SM_angular_momentum_loss}.

  At leading self-force order, this boundary contribution
  equals the full static mechanical loss in the stated frame.
  Indeed, in Eq.~(4.10) of Ref.~\cite{Riva:2023xxm},
  $D_A Y^A=0$ for spatial rotations, while the remaining
  momentum term is $\mathcal O(q^3M^2)$: the outgoing total
  spatial momentum is $\mathcal O(q^2M)$ in the incoming
  center-of-mass frame, and the memory potential is
  $\mathcal O(qM)$, with $q=\mu/M$.
  The surviving shear--memory term can use either endpoint,
  since $\operatorname{Im}[\Delta h_{lm}
  \Delta \overline{h}_{lm}]=0$.
  Nonlinear memory and radiation-reaction corrections to the
  endpoint strain likewise enter beyond the retained
  $\mathcal O(q^2M^2)$ loss. This argument applies at any
  PM--PN order within the leading-self-force approximation.

  The incoming constant-strain modes are given by
  \begin{equation}
    h^-_{lm}
    =
    4\mu(-i)^m\pi\gamma v^l\,l!\,a^0_{lm}e^{im\chi_{\mathrm{cons}}/2}\,
    {}_2F_1\!\left(
      \tfrac{l-1}{2},\tfrac l2,l+\tfrac32;v^2
    \right),
  \end{equation}
  where $\gamma=(1-v^2)^{-1/2}$, ${}_2F_1$ is the Gauss hypergeometric function, and
  $a^0_{lm}$ is the numerical harmonic coefficient given in
  Eq.~(37) of \cite{Georgoudis:2025vkk}.
  The phase $e^{im\chi_{\mathrm{cons}}/2}$ accounts for the
  rotation of the incoming velocity by
  half of the geodesic scattering angle relative to the reference direction
  $+y$ in our symmetric scattering frame.

  At the retained order, the incoming and outgoing velocities
  have equal magnitudes and differ by the geodesic deflection.
  The outgoing direction is rotated by
  $+\chi_{\mathrm{cons}}/2$ relative to the same reference,
  so our azimuthal convention gives
  \begin{equation}
    h^+_{lm}=e^{-im\chi_{\mathrm{cons}}}h^-_{lm}.
  \end{equation}
  Consequently,
  \begin{equation}
    \Delta J^{\infty,\mathrm{static}}
    =
    \frac{1}{16\pi}
    \sum_{l,m}m\,|h^-_{lm}|^2
    \sin(m\chi_{\mathrm{cons}}).
    \label{eq:SM_static_angular_momentum_sine}
  \end{equation}
  We expand the constant-strain modes in $v$ and the geodesic
  angle in PM orders, retaining all products through the
  required PM--PN truncation. The harmonic sums can
  equivalently be evaluated through angular moments of
  azimuthal derivatives of the corresponding constant-strain
  field.

  The angular-momentum loss reported in the main text is
  $\Delta J^{\infty,\mathrm{rad}}$, whereas the reported
  dissipative angle uses
  \begin{equation}
    \Delta J^\infty
    =
    \Delta J^{\infty,\mathrm{rad}}
    +
    \Delta J^{\infty,\mathrm{static}}
  \end{equation}
  in Eq.~\eqref{eq:SM_dissipative_scattering_angle}.
  All losses considered here are associated with future null
  infinity; horizon contributions are excluded.

\end{document}